\documentclass[prd,superscriptaddress,amsmath,twocolumn]{revtex4-1}
\usepackage{graphicx}
\usepackage{float}
\usepackage{color}
\usepackage{epstopdf,cancel,ulem}
\usepackage{epsf,latexsym,bbm,euscript}
\usepackage{amssymb,amsmath}
\usepackage{mathtools} 
\usepackage{times,graphics}
\usepackage{soul,xcolor}
\usepackage{epsfig}
\usepackage[T1]{fontenc}
\usepackage{booktabs,appendix}

\def\6{{\langle}}
\def\9{{\rangle}}

\newcommand{\be}{\begin{equation}}
\newcommand{\ee}{\end{equation}}
\newcommand{\ba}{\begin{eqnarray}}
\newcommand{\ea}{\end{eqnarray}}

\newcommand{\beq}{\begin{equation}}
\newcommand{\eeq}{\end{equation}}
\newcommand{\beqa}{\begin{eqnarray}}
\newcommand{\eeqa}{\end{eqnarray}}

\def\be{\begin{equation}}
\def\ee{\end{equation}}

\def\bali{\begin{align}}
\def\eni{{\end{align}}}

\def\1{{{\mathbbm 1}}}

\def\half{{\tfrac{1}{2}}}

\def\pad{{\partial}}

\def\sg{\textsl{g}}

\def\cO{\mathcal{O}}
\begin{document}

\title{Transition to light-like trajectories in thin shell dynamics}

\author{Robert B. Mann}
\affiliation{Department of Physics and Astronomy, University of Waterloo, Waterloo, Ontario, Canada}
\affiliation{Perimeter Institute for Theoretical Physics, Waterloo, Ontario, Canada}
\author{Ian Nagle}
\affiliation{Department of Physics \& Astronomy, Macquarie University, Sydney NSW 2109, Australia}

\author{Daniel R. Terno}
\affiliation{Department of Physics \& Astronomy, Macquarie University, Sydney NSW 2109, Australia}

\begin{abstract}
It was recently shown that a massive thin shell that is sandwiched between a flat interior and an exterior geometry   given by the outgoing Vaidya metric becomes null in a finite proper time. We investigate this transition for a general spherically-symmetric metric outside the shell and find that it occurs generically. Once the shell is null its persistence on a null trajectory can be ensured by several {mechanisms} that we describe. Using the outgoing Vaidya metric as an example we show that {if a dust shell acquires surface pressure on its transition to a null trajectory it can  evade} the Schwarzschild radius through its collapse.  Alternatively, the pressureless collapse may continue if the exterior geometry acquires a more general form.
\end{abstract}
\maketitle


\section{Introduction}

Hypersurfaces of discontinuity are idealizations of narrow transitional regions between  spacetime domains with different physical properties. The thin shell formalism \cite{lanczos:24, israel, poisson} makes this idealization consistent by prescribing joining rules  for the  solutions of the Einstein equations  on both sides of the hypersurface.  These rules --- junction conditions ---  determine dynamics of the shell. The resulting joined  geometry is a  solution of the  Einstein equations with an additional  distributional stress-energy tensor that is  concentrated on the hypersurface.

Thin shell formalism plays a role in  studies of cosmological phase transitions \cite{phase}, impulsive gravitational waves \cite{bh:98}, gravastars and other non-singular substitutes of black holes \cite{regular}, traversable wormholes \cite{wormhole},  and gravitationally-induced decoherence \cite{gu:15}.   A massive thin shell separating a flat interior from a
curved exterior spacetime region provides  the simplest model of  collapse.  Classically the exterior spherical geometry is described by a Schwarzschild metric and the shell collapses into a black hole in finite proper time.

 {Such models} has also been used in  investigations of  collapse-induced radiation \cite{vsk, pp:09}  anticipated
before formation of the event horizon.
The basic idea is that the process of gravitational collapse excites fields  in the spacetime, giving rise to \ {asymptotically} thermal radiation
 \cite{vsk}.  
   We shall refer to this   as  pre-Hawking radiation \cite{haj:87}.  The consequences it might have for black hole formation and the information paradox have been a subject of interest in recent years
\cite{vsk, pp:09,kmy:13,ho:16,Mersini-Houghton:2014cta,us-1,us-new,  Barcelo:2017lnx,Kawai:2017txu,Arderucio-Costa:2017etb}.
While it has been argued that such effects are too small to prevent the formation of an
event horizon \cite{pp:09,Arderucio-Costa:2017etb},  others
{contend} that such approximations are not reliable and that horizons may not form if pre-Hawking radiation is properly taken into account
\cite{ho:16,Mersini-Houghton:2014cta}.  Indeed, it has been posited that this should be a generic feature of quantum gravity, with the black hole interior and accompanying singularity   replaced with a genuine quantum geometry where the notion of event horizon ceases to be useful  \cite{Ashtekar:2005cj}.

A number of researchers have argued \cite{kmy:13,ho:16,us-1,us-new} that there are two options for the evolution of   a thin shell in a spacetime with   pre-Hawking radiation. One possibility is that an event horizon never forms:  either the
 shell does not cross its Schwarzschild radius $r_\sg$ before complete evaporation or   a manifest breakdown of  semiclassicial dynamics, such as formation of a Planck-scale remnant \cite{coy:15}, violation of the adiabatic condition \cite{blsv:11}, or
 formation of some quantum geometry \cite{Ashtekar:2005cj} occurs. The other alternative is that   evaporation stops, forever preventing a distant observer at late times
 from detecting  Hawking radiation.
 An outgoing Vaidya metric \cite{vai:51} is often used as an  example of the exterior geometry of this process despite its known limitations \cite{bmps:95}.

This result is based on an implicit assumption that through their evolution a massive shell remains timelike and a massless shell remains null. However, it was recently demonstrated by Chen, Unruh, Wu and Yeom (CUWY) \cite{cuwy:17} that this assumption is unwarranted. Indeed, it was  shown  that if the exterior metric outside is rigorously Vaidya, a massive dust shell sheds its rest mass in finite proper time (while still outside its Schwarzschild radius),  {becoming null}.  It was further argued that if the evaporation continues the shell becomes superluminal.  The choice is evidently between {eventual tachyonic behaviour} or switching off the radiation. In the latter case the subsequent development is  classical  and the shell crosses  $r_\sg$ at a finite value of a suitable affine parameter \cite{cuwy:17}. 

 Motivated by the goal of understanding the limits of validity of the semiclassical approximation in the context of gravitational collapse, we
focus here on the dynamics of thin shells in spacetimes that model  evaporation due
to pre-Hawking radiation.  The detailed description of the basic assumptions of this approximation and their application to 
thin shells is given in \cite{us-1}. In practical terms, the standard curvature terms of the left hand side of the Einstein equations are equated to the expectation value of the renormalized stress-energy tensor. We assume its existence and consistency, but make no assumptions beyond  that of spherical symmetry  and certain regularity conditions that are described below. We leave aside the conceptual  implications of radiation suppression  {and/or horizon avoidance} \cite{us-1, bmt-e}.  The question of the origin   of the pre-Hawking   quanta  is open, as it has been posited that this takes place at or near the surface  of  the collapsing body  \cite{ho:16} or within a region  $\Delta r\sim r_{\sg}$ outside the  horizon \cite{giddings:16}.

The flat geometry inside is given in the outgoing Eddington-Finkelstein (EF) coordinates,
\be
ds_-^2=-du_-^2-2du_-dr+r^2d\Omega,
\ee
where $u_-=t-r$, and the  {most general} spherically-symmetric geometry outside is
\be
ds_+^2=-e^{2h(u_+,r)}f(u_+,r)du_+^2-2e^{ h(u_+,r)}du_+dr+r^2d\Omega,  \label{genout}
\ee
 {extending the set-up of the previous studies \cite{us-1,cuwy:17}.}
In the following we omit the subscript $``+"$ from   the exterior quantities when it does not lead to  confusion. Here
\be
f(u,r)=1-C(u,r)/r,
\ee
where in the Schwarzschild geometry $C=2M=r_\sg$.

To represent   evaporation we assume that $\pad_u C\leq 0$, and the evaporation stops at some $u_*$ either at $f\equiv 1$ or with some finite value of the mass function $C(r)>0$.
The null coordinates $u_\pm$ are distinct and the relation  between them is determined by the first junction condition, while the radial coordinate is continuous across the surface \cite{poisson}.

We consider the transition of a massive evaporating shell to a null trajectory and investigate
the circumstances under which such a  shell continues along a null trajectory.  We find
this does not take place only if $h(u,r)\equiv 0$ \textit{and} the absence of surface tension or pressure is imposed  on the   shell when it becomes null. Provided that the metric at the Schwarzchild radius is regular --- specifically, the function $h\big(u,r_\sg(u)\big)$ is finite, we show by reworking the arguments of \cite{us-new} that the subsequent null trajectory will never cross the ever-shrinking $r_\sg(u)$. In this sense the
massive-to-null-to-superluminal case considered by CUWY is  exceptional.
We discuss the physical implications of the various possible cases in the concluding section of our paper, noting that which, if any, scenario is realized can be decided only by performing explicit analysis of the pre-Hawking radiation.

To simplify the notation in the following we use $w:=u_-$ and refer the {quantities} on the shell $\Sigma$ by   capital letters, such as $R:=r_{|\Sigma}$,  $F:=f(U,R)$.  The jump {of some quantity} $A$ across the shell is $[A]:=A|_{\Sigma_+}-A|_{\Sigma_-}$. All derivatives are explicitly indicated by subscripts, as in $A_R=\pad_R A(U,R)$.  The total  proper time derivative  $dA/d\tau$ is denoted as $\dot A$,  and the total derivative over some parameter $\lambda$ is $A_\lambda:=A_R R_\lambda+A_U U_\lambda$.

\section{Transition to massless shell}

The metric across the two domains can be represented as the continuous distributional tensor \cite{poisson}
\be
\bar{\sg}_{\mu\nu}=\bar{\sg}_{\mu\nu}^+{\Theta}(z)+\bar{\sg}_{\mu\nu}^-\Theta(-z),
\ee
using  the set of special coordinates $\bar x^\mu=(w,z,\theta,\phi)$. Here $\Theta(z)$ is the step function and the interior and exterior metrics $\bar \sg^\pm(\bar x)$ are continuously joined  at $z=0$.  Mathematically equivalent  and  sometimes easier to implement approach is the thin shell formalism. We will use it in most of our analysis, both for the consistency with \cite{us-1,us-new} and because it makes structure of the distributional stress-energy tensor more transparent.

{The explicit form of the interpolating metric $\bar\sg_{\mu\nu}$ when the exterior geometry that is modelled by the outgoing Vaidya metric is given in \cite{cuwy:17}. We treat a general spherically-symmetric exterior geometry and provide the  resulting metric $\bar{\sg}_{\mu\nu}$}  in Appendix~\ref{apint}.

In discussing the timelike-to-null transition it is particularly convenient to have a unified description that is applicable to both  types of shells \cite{bi:91}.  Unlike the proper time $\tau$    that diverges at the transition to the null trajectory, two additional parameterizations are  regular there. When discussing the null shell it is convenient to use
\be
\lambda:=-R, \label{lar}
\ee
while the Minkowski retarded coordinate $w$ is used in calculations that involve the interpolating metric.  {We will primarily use the thin shell formalism in $\tau$ and $\lambda$ parameterizations}.

 We first consider an initially massive thin shell $\Sigma$ and assume that the exterior geometry is described by the outgoing Vaidya metric (eq. \eqref{genout} with$f(u,r)=1-C(u)/r$, for some decreasing function $C(u)$, $h(u,r)=0$). While the shell is timelike its four-velocity is given by
 \be
 v^\mu_\pm=\dot\lambda k^\mu_\pm,
 \ee
where
\be
k^\mu_+:=(U_\lambda,-1,0,0), \qquad k^\mu_-:=(W_\lambda,-1,0,0) \label{veck},
\ee
and
\be\dot\lambda=(-k^\mu k_\mu)^{-1/2}.
\ee
The first junction condition identifies the induced metric on the two sides of $\Sigma$,
 \be
W_\lambda^2-2 W_\lambda=FU_\lambda^2-2U_\lambda,
\ee
where $F=f(U,R)$.  Similarly,  the condition $\bar k^\mu\equiv \bar k^\mu_-=\bar k^\mu_+$  (see Appendix \ref{apint})
holds both for a massive shell  and in the lightlike regime, where $k^2_\pm=0$. For a massive shell we also have
\be
\dot U=\frac{-\dot R+\sqrt{F+\dot R^2}}{F}, \label{udote}
\ee
that approximately becomes $\dot U\approx-2\dot R/F$ for large $-\dot R$, and for the null shell
\be
U_\lambda=2/F.
\ee
 An important auxiliary quantity for a massive shell is the outward pointing (unit) spacelike normal,
\be
\hat n^\mu=\dot\lambda n^\mu, \qquad n_\mu=(1, U_\lambda,0,0),
\ee
and as the shell approaches the null trajectory $n_\mu\rightarrow -k_\mu$. Both $\bar k^\mu$ and $\bar n^\mu$ are continuous across the shell (when written in the interpolating coordinates $\bar x^\mu$).

The second junction condition relates the jump in   extrinsic curvature
\be
K_{ab}:=\hat n_{\mu;\nu}e^\mu_a e^\nu_b, \label{extK}
\ee
to the surface stress-energy tensor. Here we use the surface coordinates $y^a$, $a=1,2,3$, the shell is given via parametric expressions $x^\mu_\pm(y)$, and $e^\mu_a=\pad x^\mu/\pad y^a$. In this case the optimal choise of the surface coordinates is $(\tau,\Theta:=\theta|_{\Sigma},\Phi:=\phi|_{\Sigma})$.

Assuming a general relationship  between the proper mass density  (that is related to the shell's rest mass via $m_0=4\pi R^2\sigma$) and the tension/pressure $p(\sigma)$, we obtain equations that govern their evolution,
\begin{align}
 8\pi p(\sigma)=&\frac{2\ddot R + F_R}{2\sqrt{F+\dot R^2}} - \frac{\ddot R}{\sqrt{1+\dot R^2}}+\frac{F_U\dot U^2}{2\sqrt{F+\dot R^2}} \nonumber\\
& +\frac{\sqrt{F+ \dot R^2} - \sqrt{1+\dot R^2}}{R}, \label{eom}
\end{align}
and
\be
- 4\pi\sigma = \frac{\sqrt{F+\dot R^2}-\sqrt{1+\dot R^2}}{R}. \label{tautau}
\ee

The details of the derivation can be found in \cite{us-1}. The system can be solved for $\ddot R$ providing the basis for numerical integration. In the limit of large $\dot R$ the pressure is negligible, and the asymptotic expression becomes \cite{us-1}
\be
\ddot R\approx\frac{4 C_U\dot R^4}{RF^2}\approx \frac{4CC_U\dot R^4}{X^2}, \label{ddrap}
\ee
where we defined the gap between the shell and the Schwarzschild radius,
\be
X:=R-C,
\ee
and the second equality in Eq.~\eqref{ddrap} holds for $X\ll C$.

\begin{figure}[htbp]
\includegraphics[width=0.45\textwidth]{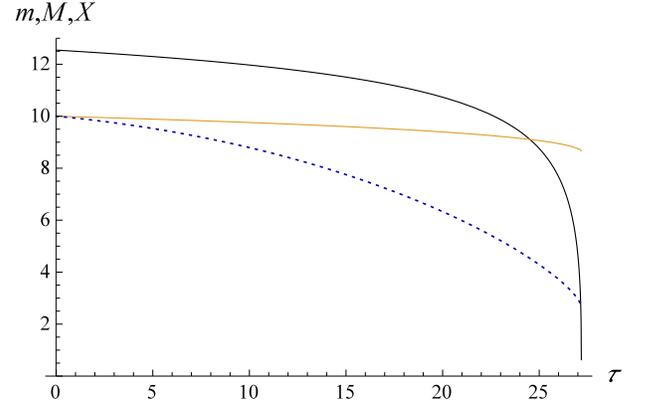}
\caption{The orange line represents $M(\tau)\equiv C(\tau)/2$. The rest mass $m_0(\tau)$ is shown as the thin black line, and the gap $X(\tau)=R(\tau)-C(\tau)$ as the blue dotted line.  The initial conditions are as in \cite{cuwy:17}, where $R_w(0)=-0.1$ is translated into $\dot R(0)=-0.111803$ using Eq.~\eqref{conver}. Other initial data is $C_0\equiv C(0)=20$   and $R(0)=30$. The  coefficient in the evaporation law  Eq.~\eqref{evlaw} is $\alpha=8$ that corresponds to $\alpha=1$ of \cite{cuwy:17}. The evaporation ends at time is $u_*=C_0^3/3\alpha=1000/3$, but the system breaks down at approximately $\tau_0=27.179869$, indicating transition to the null trajectory.  Already at $\tau_m=27.1795$  ($u_m=116.409$, $w_m=39.6697$) the timelike condition $v^\mu v_\mu=-1$ is satisfied {only} with the precision of $1.76\times 10^{-9}$, {while Eq.~\eqref{estr}} gives the    estimate of the radius of convergence  $\varrho\approx 0.00049$.  At the transition most of the gravitational mass is still contained within the shell: $C(\tau_m)/C(0)=0.867$.}
\label{gap-short}
\end{figure}

To illustrate the timelike-to-null transition we set $p(\sigma)=0$ and following CUWY adapt the  law
\be
\frac{dC}{dU}=-\frac{\alpha}{C^2}. \label{evlaw}
\ee
The results of numerical integration {of \eqref{eom} with $p=0$}
are presented in Fig.~\ref{gap-short}; using  the same initial conditions as in \cite{cuwy:17} we obtain the same result.

Henceforth  we use
\be
F_R=\frac{C}{R^2}, \qquad F_U=-\frac{C_U}{R}=+\frac{\alpha}{C^2 R}, \label{evanum}
\ee
while
\be
F_\lambda=F_R R_\lambda+F_U U_\lambda=-F_R+F_U U_\lambda.
\ee

We establish  the divergence in $\dot R$ at finite proper time by considering the Taylor series for $R$ (and thus $\dot R$) at some regular point $\tau$ and showing that its radius of convergence goes to zero as $\tau$ increases. The third and higher  derivatives
{ $R^{(r)}(\tau)$ are calculated} from the knowledge of the coordinates $R(\tau)$ and $U(\tau)$, velocity $\dot R$ and the function $C(U)$. We estimate the radius of convergence of the series
using Eqs.~\eqref{ddrap} (since for large values of $|\dot R|$ it is the dominant contribution to the derivative of $\dot R$) and \eqref{evlaw}.

 In the leading term the $(r+1)$-th derivative pulls down the exponent from $\dot R$ (which equals to $3r-2$), increases the power of $\dot R$ by $3=-1+4$ and increases the power of $x$ in the denominator by 2, when we substitute $\ddot R$. As a result, the leading term in the derivatives scales as
\be
|R^{(r+1)}|\sim \frac{|\dot R|^{(3r+1)}}{(R-C)^{2r}}\left(\frac{4\alpha}{C}\right)^{r}\prod_{n=2}^r(3n-2),
\ee
and the $r$-th coefficient in the Taylor series is
\be
|c_r|=|R^{(r)}|/r!\sim \frac{1\!2^{r}}{\Gamma(\tfrac{1}{3}) r^{-{5}/{3}}}\frac{|\dot R|^{(3r+1)}}{(R-C)^{2r}}\left(\frac{\alpha}{C}\right)^{r}
\ee
Then the radius of convergence $\varrho$ is
\be
\varrho=\big(\lim_{r\rightarrow\infty}|c_r|^{1/r}\big)^{-1}\approx\frac{C X^2}{12 \alpha |\dot R|^3} \label{estr}
\ee
which goes to zero with decreasing $C$ and $X$ and increasing  $|\dot R|$.  

On the other hand, if the shell is still timelike the gap $X$ begins to increase after reaching approximately $X\approx\epsilon_*=\alpha/C$ (\cite{us-1}; in Section IV we revisit this estimate taking into account the timelike-to-null transition). Since the acceleration is negative and increasing in absolute value the transition to a null trajectory can occur only for $X>0$, i.e. outside the Schwarzschild radius.  While the above result does not allow identification of the transition point $\tau_0$, it shows that it exists.

Lightlike matter has a vanishing   rest mass.  Eq.~\eqref{tautau} ensures that when the shell becomes null [i.e. $\dot R\rightarrow-\infty$], its surface density goes to zero,
\be
\sigma=\frac{C}{8\pi|\dot R |R^2} +\cO(1/|\dot R^3|),
\ee
causing the rest mass {$m_0=4\pi\sigma R^2$ to} vanish. The rate of shedding of the rest mass at large velocities is
\be
\dot m_0\sim -\frac{\dot C}{2\dot R}+\frac{C\ddot R}{2\dot R^2}\approx-\frac{2|C_U| C^2\dot R^2}{X^2},
\ee
 where only the term proportional to $\ddot R<0$, approximated by using Eq.~\eqref{ddrap},  appreciably contributes to the final result. We see that this rate is
 much higher than the rate of decrease of $C$.  Indeed, for macroscopic shells $C\gg 1$ the fraction of the gravitational mass lost before the transition to the null trajectory goes to zero slover than $1/C$ (Appendix \ref{closs}).

In Appendix~\ref{apdiv} we show that such transitions happen for a general spherically-symmetric exterior metric.

\section{Preserving the null condition}
For the shell becoming null at $\lambda=\lambda_0$ we now investigate the conditions necessary to keep it null in  a general spherically symmetric metric \eqref{genout}. First we recall a few properties of the surface stress-energy tensor  in the null case. Since the normal $n_\mu$ ``declines into tangency with $\Sigma$'' \cite{bi:91}  an alternative auxiliary vector is used, defined by
\be
N_\mu^\pm N^\mu_\pm=0, \qquad N_\mu^\pm k^\mu_\pm=-1
\ee
on both sides of the shell. When the shell  becomes null at some $\lambda=\lambda_0$
\be
N^\mu_+=(0,\half F,0,0), \qquad N^\mu_-=(0,\half,0,0).
\ee
Analogously to the vector $\bar k^\mu$ it is continuous at $\Sigma$, $[\bar N^\mu]=0$. 
On the null hypersurface  {orthogonal and tangent to the 4-velocity of the null shell}
(that is traversed by the shell if it remains null)
 we install the coordinates $y^a$ that consist of $\lambda\geq\lambda_0$, possibly non-affine parameter of the hypersurface generators, and the transversal $y^{A}$ that are most conveniently taken to be $(\Theta,\Phi)$.   From the bulk point of view the shell is described by parametric relations $x^\mu(y^a)$, and the set of three tangent vector fields  is formed by two spacelike vectors
\be
e^\mu_A:=\frac{\pad x^\mu_\pm}{\pad y^A}\Big|_\Sigma,
\ee
 transverse to it  that are also continuous across the surface, and the null vector
 $e^\mu_\lambda:=k^\mu$.
The transverse inner product $\sigma_{AB}:=e^\mu_A e_{B\mu}$ is continuous as well. For a spherical shell it is
\be
\sigma_{AB}dy^Ady^B=\lambda^2(d\Theta^2+\sin^2\Theta d\Phi^2).
\ee
The surface stress-energy tensor of a massless shell depends on the observer. However, all such objects are derived from
\be
S^{\mu\nu}=\varsigma k^\mu k^\nu+p \sigma^{AB}e^\mu_A e^\nu_B+j^A(k^\mu e^\nu_A+e^\mu_A k^\nu), \label{surbase}
\ee
where $\varsigma$ is the shell's surface energy density, $p$ is an isotropic surface pressure, and  $j^A$ is  surface current that is zero in the spherical case. From the extrinsic perspective the three above parameters are related to the discontinuity of the derivative of the continuous interpolating metric via $\gamma_{\mu\nu}:=[\bar g_{\mu\nu,\alpha}]\bar N^\alpha$ where the calculation is performed using the coordinates $\bar x^\mu$.  From the intrinsic perspective the key quantity is a transverse curvature $\mathfrak{C}_{ab}$,
\be
\mathfrak{C}_{ab}:=-N_\mu e^\mu_{a;\nu}e^\nu_b.
\ee

Density, pressure and current are obtained from its discontinuity, $[\mathfrak{C}_{ab}]=\half \gamma_{ab}e^\mu_a e^\nu_b$.  A straightforward calculation identifies
\be
\varsigma=\frac{C}{8\pi R^2}, \label{endens}
\ee
while the pressure is directly connected to the preservation of the null condition $k_\mu k^\mu=0$,
\be\label{pkap}
p=-\frac{[\kappa]}{8\pi},
\ee
where $\kappa=\mathfrak{C}_{\lambda\lambda}=-N_\mu k^\mu_{;\nu}k^\nu$ measures the failure of $\lambda$ to be the affine parameter, with acceleration $a^\mu_\pm:=k^\mu_{\pm;\nu}k^\nu_\pm$ being
\be
a^\mu_\pm=\kappa_\pm k^\mu_\pm, \label{affail}
\ee
on either side of the shell.  Note that unlike its massive counterpart, a massless shell moves on a geodesic, possibly non-affinely parameterized. The shell  that becomes null at $\lambda=\lambda_0$ and continues as as null for $\lambda>\lambda_0$    should satisfy Eq.~\eqref{affail}  already at $\lambda=\lambda_0$.

At this stage two options are possible. One is that the spacetime outside the shell is still described by the outgoing  Vaidya metric.  The other is that the form of the metric changes at the null transition.  We shall explore each in turn.

Suppose
the spacetime outside the shell  retains its outgoing  Vaidya form.
 The  shell will continue on a null trajectory for $\lambda>\lambda_0$, provided it acquires a surface pressure. Indeed, the acceleration of the shell expressed in the outside and the inside coordinates is
\be
a^\mu_+=\begin{pmatrix}
U_{\lambda\lambda}-{2F_R}/{F^2} \\
{2F_U}/{F^2}
\end{pmatrix}, \qquad a^\mu_-=\begin{pmatrix}
W_{\lambda\lambda}\\
0
\end{pmatrix}, \label{vaida}
\ee
respectively,  where we suppressed the trivial angular components. Then from Eqs.~\eqref{veck} and \eqref{vaida} it follows that
\be
\kappa_-=0, \qquad \kappa_+=-{2F_U}/{F^2},
\ee
resulting in  the surface pressure
\be
p=-\frac{C_U}{4\pi R(1-C/R)^2}>0 \label{presV}
\ee
upon using \eqref{pkap}.
Furthermore, (as we shall see in Sec.~\ref{horav})
 $R\geq C+\epsilon_*$, and so the pressure is finite for all finite values of $C_U$.
 The shell moves on a null geodesic with a consistent value of the second derivative $U_{\lambda\lambda}\equiv d(2/F)/d\lambda$ for $\lambda\geq\lambda_0$.

Alternatively, one could impose the requirement  $p\equiv0$ as was done by CUWY \cite{cuwy:17}. A combination of the null shell property $W_\lambda=2$ (Eq.~\eqref{wlam}) and
\be
F_\lambda=-F_R+F_UU_\lambda=-F_R+2F_U/F,
\ee
{then ensures that the only consistent  solution for $\lambda>\lambda_0$ that is not superluminal is
\be
F_U\equiv 0, \qquad U_{\lambda\lambda}=\frac{2F_R}{F^2}=-\frac{2F_\lambda}{F^2},  \qquad W_\lambda=2.
\ee
{In other words, the pre-Hawking radiation cuts off and (as $C_u$  must drop to zero
at} $\lambda=\lambda_0$) the metric has a discontinuity in the first derivative, reducing it to the Schwarzschild metric expressed in Eddington-Finkelstein coordinates.  We discuss the superluminal solution in Appendix \ref{apta}.

 We therefore see that
it is impossible to have both a pressureless null shell and  a shrinking Schwarzschild radius $dr_\sg/du<0$ if the exterior metric is the outgoing Vaidya.

The surface stress-energy tensor of Eq.~\eqref{surbase} satisfies the weak energy condition for positive values of $\varsigma$ and $p$. Indeed, for an arbitrary timelike vector $t^\mu=(\dot u, \dot r,\dot\theta,\dot\phi)$ we have
\be
S_{\mu\nu}t^\mu t^\nu=\varsigma (k_\mu t^\mu)^2+pR^2(\dot\theta^2+ {\sin^2\!\Theta}\dot\phi^2)\geq 0.
\ee

 The transverse pressure $p$, however, may not be a benign feature of the solution. Having normal matter
is not sufficient to rule out the superluminal propagation of disturbances, i.e. to guarantee that the speed of sound is less then
the speed of light \cite{cur:16}.  For the most part the shell is super-stiff, i.e.,  $p>\varsigma$.  { Note that motion of the shell determines the   surface quantities
$(p,\varsigma)$ via the junction conditions;  there is no equation of state and thus no well-defined speed of sound. }

This leads us to the other option: the metric has a different discontinuity at the null transition
allowing both zero pressure and a shrinking $r_\sg$. For a general metric \eqref{genout} the two first components of 
 Eq.~\eqref{affail} become
\begin{align}
&U_{\lambda\lambda}+\left(-\half e^HF_R-e^HFH_R+H_U\right)\frac{4 e^{-2H}}{F^2}=\kappa \frac{2e^{-H}}{F},\label{trueull}\\
& \frac{2e^{-H}F_U}{F^2}+H_R=-\kappa,  \label{capgen}
\end{align}
where $H=h(U,R)$. {Given the functions $f$ and $h$ this pair of equations yields $\kappa$ and $U_{\lambda\lambda}$. The shells continues on a null trajectory, with the velocity $U_\lambda= 2e^{-H}/F$ in a general metric of Eq.~\eqref{genout}} satisfying}
\be
 U_{\lambda\lambda}=d(2 e^{-H}/F)/d\lambda.
\ee

Since $\kappa_-\equiv 0$ it is enough to require $\kappa=0$ to ensure that the shell remains pressureless while continuing  to move on a null geodesic.
{In this case Eq.~\eqref{capgen} will serve as a constraint on the exterior metric: $H_R=-2e^{-H}F_U/F^2$}.
The evaporation continues  {with} $F_U<0$.  There is a discontinuity in the derivative of the metric, $h\big(U(\lambda_0),R(\lambda_0)\big)=0$, $\pad_Rh\big(U(\lambda_0),R(\lambda_0)\big)=-2F_U/F^2$, and for $\lambda>\lambda_0$ the metric is not of the Vaidya form,  {but rather of the form \eqref{genout}.}

The question concerning which scenario is actually realized can be answered only through the detailed studies that involve matching of the bulk stress-energy tensor that results in a self-consistent analysis of evaporation. 


\section{Horizon avoidance} \label{horav}

 We have seen that the shell, once it becomes null, continues as such.  Here we show how the event horizon is avoided if we still model the exterior geometry by the outgoing Vaidya metric.  In this case the shell must have a surface pressure \eqref{presV}. A general analysis, including comparison of the evolution described in different coordinate systems will be presented elsewhere.

The key quantity is the gap
\be
X=R-r_\sg=R-C,
\ee
where the last equality holds only for the Vaidya metric. This quantity can be viewed as either function of $w$ or $\lambda$, via the relationships $R(w)$, $U(w)$, or $R(\lambda)$, $U(\lambda)$ respectively.
Evaluating its derivative over, e.g., $\lambda$, we have
\be
X_\lambda=R_\lambda -\frac{dC}{dU}U_\lambda=-1+\left|\frac{dC}{dU}\right|\frac{2}{F}>-1+\frac{2|C_U| C}{X}.
\ee
As a result the gap decreases only until $X\approx \epsilon_*:=2C|C_U|$, and crossing of the Schwarzschild radius is possible only if the evaporation completely stops.
A detailed evaluation of the approach to $\epsilon_*$ is given in Appendix \ref{closs}.

\section{Discussion}

{If gravitational  collapse is accompanied by emission of  pre-Hawking radiation
(that does not get cut off) then
initially massive thin shells  shed their rest mass and become null. }
This is a generic property of  the semiclassical model of a  {massive spherical collapsing body as a thin shell that is} sandwiched between   flat Minkowski spacetime and a generic spherically-symmetric spacetime  that is self-consistently generated via emission of radiation.
Once the shell becomes null  there are several options, that are best illustrated by the evolution of a massive dust shell with the exterior geometry given by the outgoing Vaidya metric

\begin{enumerate}
\item The pre-Hawking radiation halts at or before the null transition. The
metric retains its
Vaidya form but has derivative discontinuity, {though this perhaps could be ameliorated if the process halts sufficiently smoothly.}
 Collapse to a black hole proceeds  classically on a null geodesic (as a hypersurface separating Minkowski and  Schwarzschild geometries). 

\item The  metric retains its  Vaidya form and the shell remains pressureless, in which case it must   become superluminal. This option can be discarded as unphysical \cite{cuwy:17}.

\item The shell acquires surface pressure discontinuously  (though one
could consider modelling this as a smooth but rapid transition)
and propagates on a null geodesic (as a hypersurface separating Minkowski and outgoing Vaidya geometries).
In this case an horizon does not form, as shown in section \ref{horav}.

\item   At the opposite extreme the  shell remains pressureless, for a part or the entire duration of its evolution. There is a derivative discontinuity in the metric, but the exterior Vaidya form
is not retained.
 The shell propagates on a null geodesic (as a hypersurface separating Minkowski and a generalized outgoing Vaidya geometries). It is possible to show by modifying the analysis of \cite{us-new} along the lines of Sec.~\ref{horav} that the shell does not cross its Schwarzschild radius.  A more plausible option is a combination of some surface pressure and $h(u,r)\neq 0$, ensuring subliminal propagation of density perturbations.

\end{enumerate}

{It is clear that pre-Hawking radiation can modify thin shell collapse in a variety of ways. {In particular, it can lead to horizon avoidance or to evaporation suppression.  Whether such avoidance or suppression are universal features can be determined only from a more explicit analysis of the coupled matter-gravity systems}.


\acknowledgments
We thank Valentina Baccetti for  discussions, Bill Unruh for posing important questions about properties of the solutions, and Dieter Zeh for suggestions and helpful comments. Advice of Eric Poisson on aspects of thin shell formalism is gratefully acknowledged. {This work was supported in part by the Natural Sciences and Engineering Research Council of Canada.}

\newpage

\onecolumngrid



\appendix

\section{Interpolating metric} \label{apint}

Extending the construction of \cite{cuwy:17} we adapt the coordinates $\bar{x}^\mu=(w,z,\theta,\phi)$, where $w:=u_-$ and
\be
r=:\left\{\begin{array}{l}
                  R(w)+z, \quad z\leq 0 \\
                  R(w)+z\,{\exp{[-h\big(U(w),R(w)\big)}]}/{U_w},\quad z>0 \end{array}\right. \label{rz}
\ee
We abbreviate $h\big(U(w),R(w)\big)$ as $H(w)$.  For  general exterior point $(u,r)$ the  coordinates $(w,z)$ are obtained by identifying $u_w\equiv U_w$. Equivalently, after finding the radial  coordinate of the shell at the moment of the retarded time $w$, one obtains the equation $R_-(w)=R_+(U)\equiv R_+(u)$ that can be solved for $u(w)$. We explicitly use the subscripts indicating the spacetime domain because even if the radial coordinate is continuous, the functional dependence on the relevant retarded time is different in each region. In addition we note that
\be
R_w:=\frac{dR_-}{du_-}=\frac{dR_+}{du_+}U_w. \label{rpmd}
\ee
  {We will need the explicit form of the Jacobian on the shell:}
\be
\frac{\pad u}{\pad w}\Big|_\Sigma=U_w, \qquad \frac{\pad u}{\pad z}\Big|_\Sigma=0, \qquad \frac{\pad r}{\pad w}\Big|_\Sigma=R_w, \qquad \frac{\pad r}{\pad z}\Big|_\Sigma=\frac{e^{-H}}{U_w}. \label{jacp}
\ee

 The first junction condition in the form
\be
1+2R_w=FU_w^2+2U_wR_w,
\ee
 and the requirement that both $u_-$ and $u_+$ increase together result in the explicit expression
\be
U_w=\frac{-R_w+\sqrt{R_w^2+F(1+2R_w)}}{F}.
\ee

 The tangent vector is continuous in the coordinates $\bar x^\mu$ across the shell. Both
\be
k^\mu_+=(U_\lambda,-1,0,0)=W_\lambda (U_w,R_w,0,0), \qquad k^\mu_-=(W_\lambda,-1,0,0)=W_\lambda (1,R_w,0,0),
\ee
{  become}
\be
\bar k^\mu=\frac{\pad \bar x^\mu}{\pad x^\alpha_\pm} k^\alpha_\pm=(W_\lambda,-R_w W_\lambda-1,0,0)=(W_\lambda,0,0,0),
\ee
{where we used   Eqs.~\eqref{rz}, \eqref{rpmd} and \eqref{jacp}}.

In these   coordinates the  metric inside the shell is written as
\be
ds_-^2=-(1+2R_w)dw^2-2dwdz+(R+z)^2d\Omega_2.
\ee
 Outside the shell we have
 \be
 du=U_wdw,
 \ee
and
\be
\qquad dr=\left[R_w-\frac{z}{e^H U_w}\left(\frac{U_{ww}}{U_w}+H_w\right)\right]dw+\frac{dz}{e^H U_w},
\ee
where
\be
H_w=h_R(U,R) R_w+h_U(U,R) U_w.
\ee
As a result the  metric is
 \begin{align}
 ds_+^2=&-\left[\bar f e^{2\bar h}U_w^2+2 e^{\bar h}U_wR_w-2ze^{\bar h-H}\left(\frac{U_{ww}}{U_w}+H_w\right)\right]\!dw^2 -2e^{\bar h-H}dwdz+r^2(w,z)d\Omega_2,
 \end{align}
 where
 \be
 \bar f=1-\frac{  C\big(u(w,z),r(w,z)\big)}{r(w,z)}, \qquad \bar h(w,z)=h\big(u(w,z),r(w,z)\big).
 \ee
While the shell is timelike the normalization $v_\mu v^\mu=-1$ implies
\be
\dot W^2(1+2R_w)=1,
\ee
resulting in
\be
\dot W=-\dot R+\sqrt{\dot R^2+1}\approx -2\dot R=2\dot\lambda, \label{conver}
\ee
where the approximate equality holds for the large values of $|\dot R|$.
 When the shell becomes null the first junction condition leads to
 \be
 W_\lambda=2, \qquad R_w=-\half,  \qquad U_w=\frac{e^{-H}}{F}, \qquad U_{ww}=-\frac{e^{-H}}{F}\left(H_w+\frac{F_w}{F}\right). \label{wlam}
\ee
The interior  metric becomes
\be
ds_-^2=-2dwdz+(R+z)^2d\Omega_2,
\ee
 and the exterior metric simplifies to
\begin{align}
ds_+^2=&-\left[\frac{\bar f}{F^2} e^{2(\bar h-H)}- \frac{e^{\bar h-H}}{F}+2ze^{\bar h-H}\frac{F_w}{F}\right]dw^2 -2e^{\bar h-H}dwdz+r^2(w,z)d\Omega_2,
\end{align}
with
\be
r(w,z)=R(w)+z F(w).
\ee

\section{Estimate of gravitational mass loss and the closest approach to the Schwarzschild radius}\label{closs}
First we show that  from the moment the evaporation becomes important (or switched-on in the numerical simulation) and until the shell loses all of its rest mass only a relatively small fraction of the Bondi-Sachs mass $C/2$ evaporates. Equivalently, the elapsed interval of the EF coordinate $u$ is much smaller then the evaporation time, $\Delta U\ll u_E$.
 If the evaporation is governed by Eq.~\eqref{evlaw} then
\be
C=\big(C_0^3-3\alpha u\big)^{1/3}, \label{lawc1}
\ee
and where the evaporation time is given  by $u_E=C_0^3/3\alpha$.

For $C_0\gg 1$ we can  assume $C=C_0$ up to the transition as the first approximation. In this case using Eqs.~\eqref{lar} and \eqref{udote} we have
\be
U_\lambda \lesssim \frac{2}{F}=\frac{2 \lambda}{\lambda+C}\approx -\frac{2 C_0}{\lambda+C_0},
\ee
{where we also assume that $X\ll C$.}

The integration from the ``initial'' $R_i$ (a quantity sharply defined in the simulation and approximately in, e.g., adiabatic approximation) to the radial coordinate $R_0$ where the shell becomes null results in
\be
\Delta U\approx 2 C \ln \frac{X_i}{X_0} \approx 2 C \ln \frac{X_i}{2C|C_U|}\approx 2 C_0 \ln \frac{  C_0X_i}{2\alpha}.
\ee
The second equality is obtained by assuming that $X_0=\epsilon_*$. Numerical simulations indicates the actual value is different by a factor 2-10, but this precision is sufficient for our estimate. Substituting  this result into Eq.~\eqref{lawc1} we find that the relative reduction of the rest mass is
\be
\frac{|\Delta C|}{C_0}=\frac{\alpha}{C_0^2}\ln\frac{  C_0X_i}{2\alpha}\ll 1.
\ee

We now provide a better estimate of the mass loss that also demonstrates how the shell radius  $R$ approaches the Schwarzschild radius. We assume that $X\ll C$, but take into account  Eq.~\eqref{lawc1}. Hence the approach of the shell to the Schwarzschild radius is governed by the equation
 \be
 \frac{dU}{d\lambda}\approx-\frac{2 C(U)}{\lambda+C(U)}\quad \Leftrightarrow \quad \frac{d\lambda}{dU}=-\frac{1}{2}-\frac{\lambda}{2C}.
 \ee
The first equation is exact for the null shell. Solution of the approximate equation is
\be
-X\equiv\lambda+C=e^{C^2/4\alpha}\big(L+\sqrt{\alpha\pi}\,\mathrm{Erf}(C/2\sqrt{\alpha})\big),
\ee
where $\mathrm{Erf}(z)$ is the error function, and $L$ is determined by the initial conditions.

 It allows us to find the minimal gap $X$ between a (massive or null) shell and its Schwarzschild radius. Setting $U(\lambda_0)=0$ and  $\lambda_0=-C_0-X_0$, and {approximating the error function of a large argument as 1},   we find that
\be
L=-X_0e^{-C^2_0/4\alpha}-\sqrt{\pi \alpha},
\ee
where we suppressed the exponentially small correction terms, and for $C_0\geq C\gg \sqrt{\alpha}$ we find
\be
X=X_0 e^{-(C_0^2-C^2)/4\alpha}+\frac{2 \alpha}{C}\approx \epsilon_*,
\ee
where the last equality holds for $C_0\gg C$, in agrement with the discussion in Section~IV. Note that the condition $C^2/\alpha\gg 1$ corresponds to the adiabatic condition of \cite{blsv:11}.

\section{Transition to a null trajectory in a general case} \label{apdiv}

For a massive shell separating the flat Minkowski interior from a generic spherically-symmetric geometry of Eq.~\eqref{genout} outside, the four-velocity satisfies
\be
\dot U=\frac{-\dot R+\sqrt{F+\dot R^2}}{\bar E F},
\ee
where $H:=h(U,R)$ and $\bar E:=\exp(H)$. The outward-pointing unit spacelike normal is now
\be
\hat n_\mu=\bar E(-\dot R,\dot U,0,0).
\ee
The equation of motion for a dust shell now reads
\be
\frac{2\ddot R+F_R}{2 \sqrt{F+\dot R^2}}+\frac{\bar E F_U \dot U^2}{2\sqrt{F+\dot R^2}}
+ {H'\sqrt{F+\dot R^2}} -\frac{\ddot R}{\sqrt{1+\dot R^2}}+\frac{\sqrt{F+\dot R^2}}{  R}-\frac{\sqrt{1+\dot R^2}}{ R}=0,
\ee
and reduces to Eq.~\eqref{eom} when $h\equiv 0$ and $C(u,r)\rightarrow C(u)$. Note that in the general case $F_R=C/R^2-\pad_R C/R$.   The surface density equation \eqref{tautau} remains unchanged.

Hence the asymptotic expression for the radial acceleration is
\be
\ddot R\approx\frac{4CC_U\dot R^4}{X^2 \bar E},
\ee
 and if the function $f$ is finite outside the shrinking Schwarzschild radius the shell becomes null at a finite proper time.

\section{Alternative expressions for the surface pressure} \label{apsur}
There are additional methods to calculate the surface pressure. One is based on Raychaudhuri's equation, giving
\be
[\kappa]\theta =  [G_{\mu\nu}k^\mu k^\nu],
\ee
where $\theta$ is expansion of the geodesic congruences. In our case $\kappa_-\equiv 0$, for an incoming spherical    null shell   $\theta=-2/R$, and the non-zero components of the Einstein tensor outside the shell are
\begin{align}
G_{uu}&=e^{2h_1}\frac{\pad_rC}{r^2}\left(1-\frac{C}{r}\right)-e^{h_1}\frac{\pad_uC}{r^2},\\
G_{ur}&=e^{h_1}\frac{\pad_r C}{r^2},\\
G_{rr}&=\frac{2\pad_r h_1(u,r)}{r}.
\end{align}

A different method is based on the direct use of the discontinuity
\be
p=-\frac{1}{16\pi}\gamma_{\mu\nu}\bar k^\mu \bar k^\nu, \qquad \gamma_{\mu\nu}:=[\bar\sg_{\mu\nu,\alpha}]\bar N^\alpha.
\ee
For the interpolating metric of Appendix A  the tangent vector at the timelike-to-null transition becomes
\be
\bar k^\mu=(2,0,0,0),
\ee
and the auxiliary null vector
\be
\bar N^{\mu}=(0,\half,0,0).
\ee
Hence
\be
p=-\frac{1}{16\pi}2\times 2\times\half \bar\sg_{00,z}^+=\frac{1}{8\pi}\left(H_R+ \frac{2F_U}{e^HF^2}\right),
\ee
in agrement with Eq.~\eqref{capgen}.

\section{Tachyons and Superluminality}\label{apta}

The appearance of tachyonic behaviour is most easily observed by expressing the equation of motion for the shell in terms of the parameter $\lambda$. While it can be done starting with the definition of $K_{ab}$ in Eq.~\eqref{extK}, it is easier to substitute
\be
\dot R=-\dot\lambda=-\frac{1}{\sqrt{U_\lambda(FU_\lambda-2)}},
\ee
and
\be
\ddot R=-\frac{d\dot\lambda}{d\lambda}\dot\lambda=\half\big(FU_\lambda^2-2 U_\lambda\big)^{-2}\big(2U_{\lambda\lambda}(FU_\lambda-1)+F_\lambda U_\lambda^2\big),
\ee
into Eq.~\eqref{eom} and set $p\equiv 0$. We move directly to the asymptotic expression Eq.~\eqref{ddrap}. In $\lambda$-parameterization it becomes
\be
2U_{\lambda\lambda}(FU_\lambda-1)+F_\lambda U_\lambda^2=\frac{8 C|C_U|}{(R-C)^2}.
\ee
At the timelike-to-null transition at $\lambda_0$ we have $U_{\lambda_0}=2/F(\lambda_0)\equiv 2/F_0$. Hence we have
\be
U_{\lambda_0\lambda_0}+\frac{2F_{\lambda_0}}{ {F^2_0}}=-\frac{4C_0|C_{U_0}|}{(R_0-C_0)^2}.
\ee
If the right hand side of this equation is non-zero, then the null consistency condition $U_{\lambda\lambda}=d(2/F_\lambda)/d\lambda$ is violated and the {shell must become tachyonic}. However, this  happens solely because the equation lacks the pressure contribution  that appears in the correct equation of motion Eq.~\eqref{trueull}.


\begin{thebibliography}{9}
\bibitem{lanczos:24} C. Lanczos,  Ann. Physik \textbf{74}, 379 (1924).
\bibitem{israel} W. Israel, Nuovo Cimento \textbf{448}, 1 (1966); \textbf{488}, 463 (1967).
\bibitem{poisson} E. Poisson, \textit{A Relativist's Toolkit}, (Cambridge University Press, Cambridge, 2004).
\bibitem{phase} S. Coleman, Phys. Rev. D \textbf{15}, 2929 (1977);  V. A. Berezin, V. A. Kuzmin, and I. I. Tkachev, Phys. Rev. D
\textbf{36}, 2119 (1987).
\bibitem{bh:98} C. Barrab\`{e}s and P. A. Hogan, Phys. Rev. D \textbf{58}, 044013 (1998).
\bibitem{regular} J. P. S. Lemos and V. T. Zanchin Phys. Rev. D \textbf{83}, 124005 (2011); P. Martin-Moruno, N. Montelongo Garcia, F. S. N. Lobo, and M. Visser, JCAP \textbf{03}, 034 (2012).
\bibitem{wormhole} V. Faraoni and W. Israel, Phys. Rev. D\textbf{ 71}, 064017 (2005).
\bibitem{gu:15} C. Gooding and W. G. Unruh, Found. Phys. \textbf{45}, 1166 (2015).
\bibitem{vsk} T. Vachaspati, D. Stojkovic, and  L. M. Kraus, {Phys. Rev. D} \textbf{76}, 024005 (2007); A. Saini and D. Stojkovic, Phys. Rev. Lett. \textbf{114}, 111301 (2015).
\bibitem{pp:09} A. Paranjape and T. Padmanbhan, Phys. Rev. D \textbf{80}, 044011 (2009).
\bibitem{haj:87} P. Hajicek, Phys. Rev. D \textbf{36}, 1065 (1987).

\bibitem{kmy:13} H. Kawai, Y. Matsuo, and Y. Yokokura,   {Int. J. Mod. Phys. A} \textbf{28}, 1350050 (2013).
\bibitem{ho:16}  P.-M. Ho, {Nucl. Phys. B} \textbf{909},   394 (2016).
\bibitem{us-1} V. Baccetti, R. B. Mann, and D. R. Terno, arXiv:1610.07839 (2016).
\bibitem{us-new} V. Baccetti, R. B. Mann, and D. R. Terno, arXiv:1703.09369 (2017).

\bibitem{Mersini-Houghton:2014cta}
  L.~Mersini-Houghton and H.~P.~Pfeiffer,
  arXiv:1409.1837.


\bibitem{Barcelo:2017lnx}
  C.~Barceló, R.~Carballo-Rubio and L.~J.~Garay,
  JHEP {\bf 1705}, 054 (2017).


 \bibitem{Kawai:2017txu}
  H.~Kawai and Y.~Yokokura,
  Universe {\bf 3},  51 (2017).




 \bibitem{Arderucio-Costa:2017etb}
  B.~Arderucio-Costa and W.~Unruh,
  arXiv:1709.00115 (2017).

 \bibitem{Ashtekar:2005cj}
  A.~Ashtekar and M.~Bojowald,
 Class.\ Quant.\ Grav.\  {\bf 22}, 3349 (2005).




\bibitem{bmt-e} V. Baccetti, R. B. Mann, and D. R. Terno, Int. J. Mod. Phys. D \textbf{26}, 1743008 (2017).
\bibitem{coy:15} P. Chen, Y. C. Ong, D.-h. Yeom, Phys. Rep. \textbf{603}, 1 (2015).
\bibitem{blsv:11} C. Barcel\'{o}, S. Liberati, S. Sonego, and M. Visser, JHEP \textbf{02}, 003 (2011).
 \bibitem{vai:51} P. C. Vaidya, Phys. Rev. \textbf{83}, 10 (1951).
\bibitem{bmps:95} R. Brout, S. Massar, R. Parentani, P. Spindel,   {Phys. Rep.} \textbf{260}, 329 (1995).

  \bibitem{cuwy:17} P. Chen, W. G. Unruh, C-H. Wu, and D.-h. Yeom, arXiv:1710.01533 (2017).

\bibitem{giddings:16} S. B. Giddings, Phys. Lett. B \textbf{754}, 39 (2016).


\bibitem{bi:91} C. Barrab\`{e}s and W. Israel, Phys. Rev. D \textbf{43}, 1129 (1991).
\bibitem{cur:16} E. Curied, \textit{A primer on energy conditions}, in \textit{Towards a theory of spacetime theories}, D. Lehmkuhl, G. Schiemann, E. Scholz (eds.), p. 43 (Birkh\"{a}user, New York, 2016).
\bibitem{relhyd} L. Rezzolla and O. Zanotti, \textit{Relativisitc hydrodynamics}, (Oxford University Press, Oxford, 2013).
\end{thebibliography}
\end{document}